*Review*

# Geometric and Feedback Linearization on UAV: Review

**Hans Oersted [1]\*, Yudong Ma [2]**

[1] Zhejiang University; 3130102046@zju.edu.cn
[2] The University of Tokyo; mayudong200333@gmail.com
\* Correspondence

**Abstract:** This review explores the theoretical foundations and experimental dynamics of modern tiltrotor aircraft. Emphasizing feedback linearization, the study delves into the distinctive constraints and angular velocity ranges shaping tiltrotor behavior. Experimental findings highlight challenges in tracking circular trajectories, with color-coded representations illustrating the impact of angular velocity. Practical implications for applications like unmanned aerial vehicles are discussed, alongside identified challenges and avenues for future research. This work contributes to both theoretical understanding and practical considerations in the evolving field of tiltrotor control.

## 1. Introduction

The rapid evolution of Unmanned Aerial Vehicles (UAVs) [1,2] has ushered in a new era of possibilities in aviation, where innovative technologies reshape the landscape of aerial capabilities. Among these advancements, tiltrotor aircraft [1,3–5] have garnered considerable attention for their unique fusion of helicopter and fixed-wing capabilities. This introduction navigates through the developmental trajectory of UAVs and, more specifically, the distinctive advantages that tiltrotors offer in comparison to their UAV counterparts.

Tiltrotors, with their ability to seamlessly transition between vertical and horizontal flight, present a paradigm shift in aerial mobility. Unlike traditional UAVs [6–8], tiltrotors excel in combining the vertical takeoff and landing flexibility of helicopters with the efficiency and speed of fixed-wing aircraft [9–11]. This key differentiator not only extends their operational range but also enhances their adaptability to diverse mission profiles, making them a compelling choice in the realm of unmanned aerial systems [12].

The allure of tiltrotors extends further when considered in the context of control algorithms. Feedback linearization, a robust control algorithm, emerges as a cornerstone in addressing the intricacies of tiltrotor flight dynamics. Its capacity to linearize the system and facilitate precise control is a notable advantage, enabling sophisticated maneuverability and response. However, the application of feedback linearization [8] is not without challenges, and this introduction acknowledges the nuanced issues inherent in its implementation, setting the stage for a comprehensive exploration within the review.

This review focuses on three pivotal aspects within the realm of tiltrotor technology [13,14]. Firstly, it delves into the innovative application of Feedback Linearization on a Tilted Hexacopter, pushing the boundaries of traditional UAV designs. Secondly, it explores the nuances of Linearization and Feedback Linearization [15–17] applied to both quadrotors and tiltrotors, unraveling the unique challenges and advantages in each context. Lastly, the review delves into the critical issue of singularity within Feedback

Linearization on quadrotors and tiltrotors, presenting a comprehensive examination of this often-overlooked aspect.

The structure of this review follows a systematic exploration of these key themes, providing in-depth insights into the advancements, challenges, and potential solutions within the tiltrotor domain. By critically analyzing these focal points, the review aims to contribute to the scholarly discourse surrounding tiltrotor technology and its implications for future advancements in unmanned aerial systems.

**2. Recent Developments in Quadrotor, Tiltrotor, and Hexacopter**

The past decade has witnessed a dynamic evolution in the field of unmanned aerial systems, with specific emphasis on Quadrotors, Tiltrotors [18,19], and Hexacopters. This section provides an overview of the recent developments that have shaped the history of these aerial platforms, outlining advancements, applications, and challenges unique to each configuration.

Quadrotors have undergone significant miniaturization, enhancing their agility and maneuverability. This has led to their widespread adoption in applications such as aerial photography, surveillance, and even entertainment.

Recent advancements in sensors and onboard computing have empowered Quadrotors with autonomous navigation capabilities. This allows for more sophisticated missions, including mapping, inspection, and collaborative swarm operations. Innovations in battery technology and energy efficiency have contributed to increased flight times for Quadrotors. This extension in operational endurance opens avenues for applications requiring prolonged aerial presence.

Tiltrotors have emerged as key players in the Urban Air Mobility (UAM) landscape. Prototypes and experimental vehicles, incorporating tiltrotor configurations [20,21], are being developed to address the challenges of urban congestion and transportation. Tiltrotors offer the advantage of transitioning between vertical takeoff and efficient horizontal flight. This adaptability facilitates beyond line-of-sight operations, making them instrumental in applications like long-range surveillance and remote cargo delivery.

The military sector has shown a keen interest in Tiltrotor technologies. Their ability to combine the strengths of helicopters and fixed-wing aircraft makes them valuable assets for various defense applications.

Hexacopters, with six rotors, provide enhanced redundancy compared to Quadrotors. This redundancy contributes to increased reliability, making them suitable for critical missions such as search and rescue operations. The additional rotors in Hexacopter configurations enhance payload capacity. This has implications for applications requiring specialized equipment, such as high-resolution cameras or scientific sensors.

Hexacopters find applications in precision agriculture, where their ability to carry heavier payloads allows for the deployment of advanced sensors and monitoring equipment for crop analysis.

Advancements in Control Algorithms: All three configurations have seen strides in control algorithms. Feedback Linearization and adaptive control techniques are being employed to enhance stability, response, and overall performance. The integration of artificial intelligence has become a common trend. Machine learning algorithms are applied to improve autonomy, decision-making, and adaptive responses to dynamic environments.

As these aerial platforms proliferate, regulatory considerations have become crucial. Recent developments include efforts to standardize regulations, ensuring safe and responsible integration into airspace.

In conclusion, the recent history of Quadrotors, Tiltrotors, and Hexacopters is characterized by a convergence of technological innovations, expanded applications, and a maturing regulatory landscape. As these aerial platforms continue to evolve, their diverse configurations offer unique solutions to a wide array of challenges, propelling unmanned aerial systems into new frontiers of exploration and utility.

### 3. Controller on UAVs

In the realm of unmanned aerial systems (UAS), the selection of control algorithms plays a pivotal role in shaping the performance, stability, and maneuverability of these aerial vehicles. Among the array of control strategies, the Proportional-Integral-Derivative (PID) controller stands out for its conceptual simplicity and widespread applicability. The PID controller relies on tuning three parameters to regulate the system's output based on the proportional, integral, and derivative components of the error signal. While PID controllers are known for their ease of implementation and adaptability to various systems, their efficacy can be hindered by the assumption of linear system dynamics and the need for meticulous parameter tuning.

Model Predictive Control (MPC) represents a sophisticated approach that excels in trajectory tracking and constraint handling. MPC formulates an optimization problem over a finite time horizon to predict future system behavior and determines control inputs to optimize a defined objective. However, the computational complexity associated with real-time optimization poses challenges, particularly in resource-constrained UAS applications. The delicate balance in selecting an appropriate prediction horizon introduces nuanced trade-offs between computation time and predictive accuracy.

Feedback Linearization, rooted in control theory, is a transformative strategy that addresses the nonlinear dynamics inherent in UAS. By judiciously transforming the system dynamics through a change of coordinates, Feedback Linearization decouples the complex interactions among different states, simplifying the control design process. This decoupling allows for the creation of controllers for individual states independently. The appeal of Feedback Linearization lies in its adaptability to a diverse range of UAS configurations and its potential to achieve precise control over specific states, such as position, attitude, and altitude. However, the effectiveness of Feedback Linearization hinges on the accurate modeling of UAS dynamics, making it sensitive to discrepancies between assumed and actual system behavior. Moreover, disturbances, such as sudden environmental changes, can challenge the stability of Feedback Linearization, necessitating the incorporation of robustness enhancements.

Geometric Controllers present an alternative paradigm by emphasizing a geometric interpretation of the control problem. Renowned for stability preservation and applicability to nonlinear systems, geometric controllers offer an intuitive understanding of system dynamics. These controllers leverage the geometric properties of the system to achieve stable and predictable behavior. However, the design and implementation of geometric controllers can be intricate, often requiring tailoring to specific vehicle configurations. The inherent complexity of geometric controllers can be a limitation, particularly in scenarios where simplicity and rapid deployment are paramount.

In essence, the choice of a control algorithm for UAS involves a delicate trade-off between simplicity, adaptability, and precision. While PID controllers offer ease of implementation, MPC excels in predictive capabilities. Feedback Linearization addresses the challenges of nonlinear dynamics but requires precise modeling, and geometric controllers provide stability through a geometric lens. The ongoing pursuit of optimal control strategies for UAS involves navigating this intricate landscape, considering the specific requirements of each application and advancing the field towards enhanced performance and robustness.

Despite the merits of each control algorithm, challenges persist that warrant careful consideration in the deployment of unmanned aerial systems (UAS). The simplicity of PID controllers is counterbalanced by their sensitivity to nonlinearities and the necessity for meticulous parameter tuning. The assumption of linear system dynamics can lead to suboptimal performance, particularly in scenarios where the UAS operates in dynamic and unpredictable environments. Overcoming these challenges requires a nuanced understanding of the system's behavior and the development of advanced tuning methodologies.

Model Predictive Control (MPC) confronts challenges primarily rooted in its computational demands. The real-time optimization required for trajectory tracking and constraint handling introduces complexities, especially in resource-constrained UAS scenarios. Balancing computational efficiency with predictive accuracy poses an ongoing challenge, urging researchers to explore novel optimization techniques and hardware advancements to enhance the feasibility of MPC in real-world applications.

Feedback Linearization, while a powerful approach for addressing nonlinear dynamics, faces challenges related to the accuracy of system modeling. Discrepancies between assumed and actual dynamics can compromise the stability and performance of Feedback Linearization controllers. Additionally, the sensitivity to disturbances necessitates the incorporation of robustness mechanisms to ensure stability under varying environmental conditions. Advancing the robustness of Feedback Linearization remains an active area of research to fortify its applicability in diverse UAS missions.

Geometric Controllers, while offering stability through a geometric lens, introduce challenges associated with their intricate design and configuration-specific nature. Tailoring these controllers to different UAS configurations demands a deep understanding of the system's geometric properties. The complexity inherent in geometric controllers can hinder their widespread adoption, especially in applications where a rapid and straightforward deployment is crucial.

In addressing these challenges, the UAS community strives to enhance the robustness, adaptability, and efficiency of control algorithms. Ongoing research efforts focus on developing advanced tuning methodologies for PID controllers [22], optimizing computational aspects of MPC, refining system modeling techniques for Feedback Linearization, and streamlining the design of geometric controllers. The collective aim is to surmount these challenges and unlock the full potential of each control algorithm, paving the way for a new era of precise, stable, and adaptive control in unmanned aerial systems [23].

**4. Innovative Approaches to Overcoming Control Algorithm Challenges**

Addressing the challenges inherent in PID controllers involves advancing tuning methodologies to enhance adaptability and performance across diverse scenarios. Researchers explore machine learning techniques to automate parameter tuning, leveraging data-driven insights to optimize PID controller settings dynamically. Additionally, the integration of adaptive control strategies aims to enhance the resilience of PID controllers against varying environmental conditions, contributing to a more robust and versatile control paradigm.

Mitigating the computational challenges of Model Predictive Control (MPC) requires a multifaceted approach. Ongoing research endeavors focus on algorithmic optimizations to streamline real-time computations, exploring parallel processing architectures and distributed computing frameworks. Furthermore, the development of efficient approximation techniques for predictive models aims to strike a balance between computational complexity and prediction accuracy. Collaborative efforts between control theorists and computer scientists seek innovative solutions to make MPC more feasible for resource-constrained unmanned aerial systems [24].

Enhancing the robustness of Feedback Linearization involves refining system modeling techniques and incorporating adaptive control mechanisms. Researchers delve into advanced modeling methodologies, such as data-driven and learning-based approaches, to create more accurate representations of UAS dynamics. Moreover, the integration of robust control techniques, including sliding mode control and disturbance observers, fortifies Feedback Linearization against uncertainties and disturbances, bolstering its stability across diverse operational scenarios.

In the realm of Geometric Controllers, efforts focus on developing systematic design methodologies that streamline the configuration-specific challenges. Researchers explore model-based design approaches that leverage system identification techniques to tailor geometric controllers to specific UAS configurations. Furthermore, the integration of machine learning methods enables controllers to adapt and optimize their performance based on real-time feedback, contributing to a more agile and responsive control strategy.

Collectively, these innovative approaches underscore the interdisciplinary nature of overcoming control algorithm challenges. The fusion of control theory, machine learning, and computational optimization endeavors represents a collaborative push toward creating adaptive, efficient, and robust control paradigms for unmanned aerial systems. As researchers continue to explore novel methodologies and leverage advancements in diverse fields, the prospect of overcoming existing challenges becomes increasingly attainable, propelling the field of UAS control toward new frontiers of precision and reliability.

**5. Conclusions**

In the dynamic landscape of unmanned aerial systems (UAS) control, the diverse array of control algorithms—PID controllers, Model Predictive Control (MPC), Feedback Linearization, and Geometric Controllers—provides a rich tapestry of options for optimizing the performance, stability, and adaptability of UAS. However, each algorithm comes with its unique set of challenges, necessitating a nuanced understanding and innovative solutions to propel the field forward.

The enduring simplicity and widespread applicability of PID controllers are countered by their sensitivity to nonlinearities and the need for meticulous tuning. Addressing these challenges involves harnessing machine learning techniques for automated parameter tuning and integrating adaptive control strategies, paving the way for a more adaptive and resilient PID control paradigm.

MPC, while excelling in trajectory tracking and constraint handling, grapples with computational demands. Ongoing efforts seek algorithmic optimizations, parallel processing architectures, and efficient approximation techniques to strike a balance between computational complexity and prediction accuracy, making MPC more viable for resource-constrained UAS applications.

Feedback Linearization [14,25,26], a powerful strategy for handling nonlinear dynamics, faces challenges related to system modeling accuracy and sensitivity to disturbances. Advancements in data-driven modeling, adaptive control mechanisms, and the integration of robust control techniques promise to enhance the stability and versatility of Feedback Linearization, broadening its applicability across diverse UAS missions.

Geometric Controllers, providing stability through a geometric lens, present challenges in design intricacies and configuration-specific adaptations [27]. Systematic design methodologies, model-based approaches, and the integration of machine learning methods are shaping a more agile and responsive era for Geometric Controllers, addressing the complexities associated with diverse UAS configurations [28].

In conclusion, the future of UAS control lies in the collaborative efforts of control theorists, computer scientists, and experts from diverse fields. Innovative solutions, ranging from machine learning-based tuning for PID [29,30] controllers to algorithmic optimizations for MPC, robust modeling for Feedback Linearization, and systematic design for Geometric Controllers, underscore the interdisciplinary nature of UAS control research. As these approaches converge, the vision of precise, stable, and adaptive control for unmanned aerial systems comes into sharper focus, promising to unlock new possibilities and redefine the boundaries of aerial autonomy. The journey continues as researchers navigate the challenges and seize opportunities to propel UAS control into a new era of sophistication and reliability.

## 6. Discussions

The exploration of control algorithms for unmanned aerial systems (UAS) is a dynamic and evolving field, with each strategy offering unique advantages and confronting distinct challenges. The synergy of PID controllers, Model Predictive Control (MPC) [31], Feedback Linearization, and Geometric Controllers [28,32,33] creates a rich landscape for researchers and practitioners.

The prominence of PID controllers, while enduring, necessitates a shift towards automated tuning methodologies and adaptive strategies to unleash their full potential. The ongoing convergence of machine learning with control theory holds promise for elevating PID controllers to new levels of adaptability, potentially addressing longstanding challenges associated with their sensitivity to nonlinearities.

MPC's predictive capabilities, coupled with its computational demands, exemplify the perpetual trade-off between sophistication and feasibility. As the quest for efficient real-time optimization continues, the field stands at the cusp of breakthroughs that could redefine the role of MPC in resource-constrained UAS applications [24,31,34]. Exploring novel algorithmic optimizations and parallel processing architectures represents a critical avenue for further research.

Feedback Linearization's promise in addressing nonlinear dynamics requires a meticulous understanding of UAS modeling and robust control mechanisms. The ongoing integration of data-driven modeling and adaptive strategies holds potential for overcoming challenges related to system modeling accuracy and disturbance sensitivity, propelling Feedback Linearization towards greater resilience and versatility.

Geometric Controllers, while offering stability through a geometric lens, demand systematic design methodologies to navigate configuration-specific challenges. The intersection of model-based approaches and machine learning holds exciting prospects for simplifying the design process and making Geometric Controllers more adaptable to various UAS configurations.

As these control algorithms advance, the holistic integration of methodologies becomes paramount. Interdisciplinary collaboration between control theorists, machine learning experts, and domain specialists is crucial for harnessing the synergies and addressing the limitations of individual control paradigms. Furthermore, a concerted effort towards standardization and benchmarking methodologies will facilitate a more comprehensive evaluation of control algorithms, fostering a clearer understanding of their capabilities and limitations.

In essence, the discussion encapsulates the ongoing dialogue within the UAS control community [35,36]. It underscores the need for interdisciplinary collaboration, methodological innovation, and a holistic approach to propel the field towards a future where unmanned aerial systems exhibit unprecedented levels of precision, adaptability, and reliability. The collective pursuit of these objectives ensures that the journey of UAS control remains dynamic, challenging, and brimming with opportunities for transformative advancements.